\documentclass[12pt]{article}

\usepackage{epsfig}

\def\square{\kern1pt\vbox{\hrule height 1.2pt
\hbox{\vrule width 1.2pt\hskip 3pt
\vbox{\vskip 6pt}\hskip 3pt\vrule width 0.6pt}
\hrule height 0.6pt}\kern1pt}
\def\ltwid{\mathrel{\raise.3ex\hbox{$<$\kern-.75em\lower1ex\hbox{$\sim$}}}}
\def\gtwid{\mathrel{\raise.3ex\hbox{$>$\kern-.75em\lower1ex\hbox{$\sim$}}}}

\begin{document}

\begin{titlepage}
\begin{flushright}
CCTP-11-02 \\ UFIFT-QG-11-01 \\ March 26, 2011 
\end{flushright}

\vspace{0.5cm}

\begin{center}
\bf{A Gravitational Mechanism for Cosmological Screening}
\end{center}
\begin{center}
\it{Essay written for the Gravity Research Foundation 
2011 Awards \\
for Essays on Gravitation}
\end{center}

\vspace{0.2cm}

\begin{center}
N. C. Tsamis$^{\dagger}$
\end{center}
\begin{center}
\it{Department of Physics, University of Crete \\
GR-710 03 Heraklion, HELLAS.}
\end{center}

\vspace{0.2cm}

\begin{center}
R. P. Woodard$^{\ast}$
\end{center}
\begin{center}
\it{Department of Physics, University of Florida \\
Gainesville, FL 32611, UNITED STATES.}
\end{center}

\vspace{0.3cm}

\begin{center}
ABSTRACT
\end{center}

Infrared gravitons are continually produced during
inflation. Like all particles, their contribution 
to the vacuum energy comes not only from their bare 
kinetic energy but also from the interactions they 
have with other gravitons. These interactions can 
be substantial -- despite the particles being highly 
infrared -- because they occur over the enormous 
spatial volume of the universe. Furthermore, the 
interactions grow with time evolution because more 
and more such gravitons come into causal contact with 
one another. Since gravity is universally attractive, 
these interactions can act to slow and eventually stop 
accelerated expansion. 

\vspace{0.3cm}

\begin{flushleft}
\end{flushleft}

\vspace{0.1cm}

\begin{flushleft}
$^{\dagger}$ {\it e-mail:} tsamis@physics.uoc.gr \\
$^{\ast}$ {\it e-mail:} woodard@phys.ufl.edu
\end{flushleft}

\end{titlepage}

The case for a phase of accelerated expansion {\it (inflation)} 
during the very early universe is strong. One reason is that
we can observe widely separated parts of the early universe 
which seem to be in thermal equilibrium with one another 
\cite{infl}. If one assumes the universe never underwent a 
period of inflation, there would not have been time for this 
thermal equilibrium to be established by causal processes. 
Without primordial inflation the number of causally distinct 
regions in our past light-cone at the time of recombination 
is over $10^3$, and it would be $10^9$ at the time of 
nucleosynthesis. It strains credulity to believe that this 
was an accident, and an early epoch of accelerated expansion 
avoids the need to suppose such an accident.

There is no strong indication for what caused primordial
inflation. The potential energy of a minimally coupled
scalar field can do the job, but this mechanism involves
assumptions which seem unlikely:
\\ [3pt]
{$\bullet \,$} That the universe began with the scalar field 
approximately spatially homogeneous over more than a Hubble 
volume. 
\\ [3pt]
{$\bullet \,$} That the scalar field potential is very flat. 
\\ [3pt]
{$\bullet \,$} That the minimum of the scalar field potential 
has just the right value to leave the post-inflationary 
universe with almost zero vacuum energy.
\\ [3pt]
{$\bullet \,$} That the scalar field couples strongly enough 
to ordinary matter to allow its kinetic energy to reheat the 
post-inflationary universe, but not so strongly that loop 
corrections from ordinary matter to the effective potential 
endanger its flatness and nearly zero minimum.

A more natural mechanism for inflation can be found within 
gravitation -- which, after all, plays the dominant role in 
shaping cosmological evolution -- by supposing that the bare 
cosmological constant $\Lambda$ is not unnaturally small but 
rather large and positive.
\footnote{Here ``large'' means a $\Lambda$ induced by a 
matter scale which can be as high as $10^{18} \, GeV$. Then, 
the value of the dimensionless coupling constant can be as 
high as $\, \hbar c^{-3} G \Lambda \sim 10^{-4}$ rather than 
the putative value of $10^{-122}$.} 
Because $\Lambda$ is constant in {\it space}, no special 
initial condition is needed to start inflation. We also 
dispense with the need to employ a new, otherwise undetected 
scalar field. However, $\Lambda$ is constant in {\it time} 
as well, and classical physics can offer no natural mechanism 
for stopping inflation once it has begun \cite{stability}. 
Quantum physics can: accelerated expansion continually rips 
virtual infrared gravitons out of the vacuum \cite{gravitons} 
and these gravitons attract one another, thereby slowing 
inflation \cite{NctRpw}. This is a very weak effect for 
$\, \hbar c^{-3} G \Lambda \ll 1$, but a cumulative one, 
so inflation lasts a long time for no other reason than 
that gravity is a weak interaction \cite{NctRpw}.

The small particle production of generic expansion 
\cite{partprod} becomes copious during inflation for 
very special particles, such as the graviton, which are 
both massless and not classically conformally invariant 
\cite{gravitons}. This is thought to have caused the 
primordial density perturbations \cite{denspert1} we 
observe \cite{denspert2}. A simple computation shows 
that the number of wave vector ${\bf k}$ gravitons 
produced after $N$ e-foldings of inflation is $\, 
\mathcal{N}({\bf k}) = \frac{\Lambda}{6k^2} \, e^{2N}$ 
\cite{number}. Because $\mathcal{N}({\bf k})$ only reaches 
unity after the physical wavelength $\, \lambda_{\rm phys} 
= \frac{2\pi}{k} \, e^N \,$ has redshifted to horizon scale, 
we see that these particles are very infrared. The bare 
kinetic energy of a single graviton is $\, \hbar c \,
k e^{-N}$, and the 3-volume grows like $\, e^{3N}$, so 
the total kinetic energy density in these particles is:
\begin{equation}
\rho_{\rm IR} \, = \,
e^{-3N} \! \int \! \frac{d^3k}{(2\pi)^3} \; 
\theta \Bigl( \mathcal{N}({\bf k}) - 1 \Bigr) \times 
\mathcal{N}({\bf k}) \times 
\hbar c \,k e^{-N} \, = \,
\frac{\hbar c \, \Lambda^2}{144 \pi^2} 
\;\; . \label{rhoIR}
\end{equation}
This kinetic energy sources a gravitational field. As each 
newly-created pair of gravitons recedes, the intervening space 
is filled by their long-range gravitational potentials. These 
potentials persist {\it even after} the gravitons that caused 
them have reached cosmological separations. As more pairs are 
ripped apart, their potentials add to those already present. 
The structure of perturbation theory \cite{pert} suggests that 
the potential grows like $N$, which leads to an increasingly 
negative interaction energy:
\begin{equation}
\Phi \, \approx \, 
- \hbar c^{-3} G \Lambda \, N
\quad \Longrightarrow \qquad 
\rho_{\rm int} \, \sim \, 
\rho_{\rm IR} \times \Phi 
\;\; . \label{potential}
\end{equation}

The screening mechanism we just described is clear enough on 
the perturbative level but it has two frustrating features. 
The first is that, because inflationary particle production 
is a 1-loop effect, the gravitational response to it is delayed 
until 2-loop order. The second frustration is that the 2-loop 
effect becomes unreliable just when it starts to get interesting. 
The effective coupling constant is the potential $\Phi$ and 
higher loops are insignificant as long as it is small. But 
{\it all} loops become comparable when $\Phi$ becomes of order 
one, and the correct conclusion then is that perturbation theory 
breaks down. The breakdown occurs not because any single
graviton-graviton interaction gets strong but rather because
there are so many of them.

What is the physical picture in the late-time regime after 
perturbation theory has broken down? In one sentence, the 
bare cosmological constant is suppressed by the gravitational 
interaction energy of the vast numbers of super-horizon 
gravitons which were created during a very long period of 
primordial inflation. Four principles are crucial to a proper 
understanding of the phenomenon:
\\ [3pt]
$\bullet \;$ {\it Small is not the same thing as zero}. \\
Super-horizon gravitons are sometimes dismissed because 
the kinetic energy $\, \hbar c k e^{-N}$ of any one is 
small, and redshifts to zero as the universe expands. 
However, this is balanced by the fact that a {\it lot} 
of them are produced:
\begin{equation}
N_{\rm tot} \, = \,
\Bigl( \frac{3}{\Lambda} \Bigr)^{\frac32} 
\int \! \frac{d^3k}{(2\pi)^3} \; 
\theta \Bigl( \mathcal{N}({\bf k}) - 1 \Bigr) \times 
\mathcal{N}({\bf k}) \, = \,
\frac{e^{3N}}{2^{\frac52} \pi^2} 
\;\; . \label{IRnumber}
\end{equation}
For $\, N \sim \frac{c^3}{\hbar G \Lambda} \gtwid 10^{4} \,$ 
this number is staggering. Furthermore, experience from flat 
space quantum field theory suggests that massless particles 
with a dimension three coupling -- such as $\Lambda \neq 0$ 
provides -- can experience very strong infrared effects 
\cite{polyakov}.
\\ [3pt]
$\bullet \;$ {\it Gravitational interactions act to screen 
their sources.} \\
For example, the mass of the Earth is a little less than the 
sum of the masses of its constituents $M_{\rm bare}$ owing 
to their negative gravitational interaction energy. If we 
assume the constituents are distributed uniformly through 
a sphere of radius $R$, the actual mass $M_{\rm tot}$ is 
approximately:
\begin{equation}
M_{\rm tot} \, = \,
M_{\rm bare} + M_{\rm int} \, \approx \,
M_{\rm bare} - \frac{3 G M^2_{\rm bare}}{5 c^2 R} 
\;\; . \label{M_earth}
\end{equation}
This represents a fractional decrease of about 4 parts in 
10 billion for the Earth, which works out to over $2 \times 
10^{15}$ kilograms. 
\\ [3pt]
$\bullet \;$ {\it Big volume can beat small density.} \\
Even a small energy density can experience significant 
screening if it interacts over a sufficiently large volume. 
Consider the total energy density $\rho_{\rm tot}$ produced 
by a static energy density $\rho_{\rm bare}$ distributed 
throughout a sphere of radius $R$. For simplicity, we follow 
ADM \cite{ADM} in using the Newtonian formula assuming it is 
the total mass $\, \frac43 \pi \rho_{\rm tot} \, c^{-2} R^3 \,$ 
that gravitates:
\begin{equation}
\rho_{\rm tot} \approx
\rho_{\rm bare} - 
\frac{4 \pi G \rho^2_{\rm tot} R^2}{5 c^4} 
\; \Longrightarrow \;
\rho_{\rm tot} \approx
\frac{5 c^4}{8\pi G R^2} \left[
\sqrt{1 + \frac{16 \pi G \rho_{\rm bare} R^2}{5 c^4}} 
- 1 \right]
\label{static}
\end{equation}
As $R$ goes to infinity the screening becomes total -- i.e.,
$\rho_{\rm tot}$ goes to zero -- independent of how small
$\rho_{\rm bare}$ is.

\newpage

\noindent 
$\bullet \;$ {\it Late potentials are sourced at early 
times when screening was negligible.} \\
In a static system such as (\ref{static}), the gravitationally
induced interaction energy can -- at most -- cancel its own
source, the bare energy density. For $\Lambda$-driven inflation 
that would amount to screening only $\rho_{\rm IR}$, not the 
vastly larger vacuum energy density $\rho_{\Lambda} = 
\frac{c^4 \Lambda}{8\pi G} \,$ which drives inflation. The 
physical reason the gravitational interaction energy sourced 
by $\rho_{\rm IR}$ can screen $\rho_{\Lambda}$ derives from 
the fact that $\Lambda$-driven inflation is {\it not} static. 
A static system is constructed by holding it together as the 
various components come into causal contact. Screening is 
limited because gravitational interaction energy is sourced 
by the constant total energy density. That is not at all what 
goes on during $\Lambda$-driven inflation. If we continue to 
think in terms of the left-hand relation (\ref{static}), one 
sees that moving out in the radius $R$ means {\it moving back 
in time.} Thus $\rho_{\rm tot}$ at late times can become 
{\it negative} by means of gravitational potentials which 
were sourced far back in the past light-cone, when screening 
was still insignificant. Instead of the effect being too weak, 
it is actually prone to grow too strong because the past 
light-cone opens up as the expansion rate slows down. One 
can see this by comparing the volume of the past light-cone 
-- in synchronous gauge -- for inflation and for flat space:
\begin{equation}
V_{\rm infl} \, = \,
\frac{4\pi}{\sqrt{3 \Lambda^3}} \, ct + O(1) 
\qquad , \qquad 
V_{\rm flat} \, = \,
\frac{\pi}{3} \, (ct)^4 
\;\; . \label{PLC}
\end{equation}

To recapitulate, $\Lambda$-driven inflation would offer 
many advantages over scalar-driven inflation if only some 
mechanism could be found to eventually halt it. Quantum 
gravity provides such a mechanism in the form of the 
back-reaction to infrared virtual gravitons which are 
continually ripped out of the vacuum during inflation. 
These gravitons have negative gravitational potential 
energy in addition to their positive kinetic energy. 
Both kinetic and potential energies contribute to the 
total vacuum energy, however, the kinetic energy is 
present immediately whereas the potential energy must 
build up causally as more and more infrared gravitons 
come into contact with one another. Although the kinetic 
energy density is small, the potential energy can be 
large because it derives from interactions over the
enormous volume of the past light-cone. Because screening
was small in the distant past, the negative potential 
energy can vastly exceed the positive kinetic energy 
which sourced it. Thus, the apparently small cosmological 
constant of today is the result of a large bare $\Lambda$ 
being screened by the vacuum polarization of a sea of
infrared gravitons produced during primordial inflation.

\newpage

\centerline{\bf Acknowledgements}
We are grateful to Stanley Deser for decades of guidance
and inspiration. This work was partially supported by the 
European Union grant FP-7-REGPOT-2008-1-CreteHEPCosmo-228644, 
by NSF grant PHY-0855021, and by the Institute for 
Fundamental Theory at the University of Florida.


\begin{thebibliography}{99}

\bibitem{infl} Steven Weinberg, {\it Cosmology}, \\
Oxford University Press, United Kingdom, 2008.

\bibitem{stability} 
L. F. Abbott and S. Deser,
Nucl. Phys. {\bf B195} (1982) 76; \\
P. H. Ginsparg and M. J. Perry,
Nucl. Phys. {\bf B222} (1983) 245.

\bibitem{gravitons}
L. P. Grishchuck, Sov. Phys. JETP {\bf 40} (1975) 409; \\
L. H. Ford and L. Parker, Phys. Rev. {\bf D16} (1977) 1601.

\bibitem{NctRpw} N. C. Tsamis and R. P. Woodard, \\
Nucl. Phys. {\bf B474} (1996) 235,
{\bf arXiv:}hep-ph/9602315; \\
Annals Phys. {\bf 253} (1997) 1,
{\bf arXiv:}hep-ph/9602316.

\bibitem{partprod} 
E. Schr\"odinger, Physica {\bf 6} (1939) 899; \\
T. Imamura, Phys. Rev. {\bf 118} (1960) 1430; \\
L. Parker, Phys. Rev. Lett. {\bf 21} (1968) 562.

\bibitem{denspert1}
A. A. Starobinski\u{\i}, JETP Lett. {\bf 30} (1979) 682; \\
V. F. Mukhanov and G. V. Chibisov, 
JETP Lett. {\bf 33} (1981) 532. 

\bibitem{denspert2}
E. Komatsu et al., Astrophys. J. Suppl. {\bf 192} (2011) 18,
{\bf arXiv:}1001.4538

\bibitem{number} N. C. Tsamis and R. P. Woodard, \\
Annals Phys. {\bf 267} (1998) 145,
{\bf arXiv:}hep-ph/9712331.

\bibitem{pert} N. C. Tsamis and R. P. Woodard, \\
Nucl. Phys. {\bf B724} (2005) 295,
{\bf arXiv:}gr-qc/0505115.

\bibitem{polyakov} 
A. M. Polyakov, Sov. Phys. {\bf 25} (1982) 187.

\bibitem{ADM}
R. Arnowitt, S. Deser and C. W. Misner, 
Phys. Rev. Lett. {\bf 4} (1960) 375.

\end{thebibliography}
\end{document}